# Depth, relaxation and temperature-dependence of defect complexes in scanning transmission electron microscopy

Thomas Aarholt*[a], Ymir Frodason [a] and Øystein Prytz [a]

[a] Department of Physics, Centre for Materials Science and Nanotechnology, University of Oslo, P. O. Box 1048 Blindern, N-0316 Oslo, Norway

* Corresponding Author: thomasaarholt@gmail.com

## 1. Introduction

Doping of zinc oxide with donor atoms such as aluminium and gallium is common practice in realizing transparent conductive oxides (TCO) (1–4). For high levels of doping, formation of compensating donor-$V_{Zn}$ complexes has been found to limit the *n*-type doping efficiency and thus the performance of ZnO as a TCO (5). Understanding the exact optoelectronic interactions of the defect complexes upon the bulk is important to further improve such devices. Bandgap measurements by STEM-EELS is a method that can suitably describe such interactions. However, the first step to measure the EEL signature of these regions is to accurately recognize them in the STEM image.

Three-dimensional imaging of single vacancies in extremely thin samples has been shown possible by HAADF STEM simulation (6). Recent progress in CPU- and GPU-accelerated STEM algorithms (7–9) and user-friendly simulation software (10–12) has made it significantly easier to perform such simulations. The limits of detecting single dopant atoms have been heavily discussed in the literature. Mittal et al. (13) discussed ADF visibility of a number of dopants over a series of sample thicknesses, pointing out how a Sn defect at the top and bottom of 3 nm thick Si samples were indistinguishable. Primary-beam electron hopping between columns due to channelling and scattering has been shown to be a problem in determining the absolute composition of particles imaged by HAADF-STEM (9). Therefore, it is essential to combine simulation and experiment to determine exact compositions. With regards to imaging single defects or defect complexes within a perfect crystal, the problem is slightly simplified. Since the neighbouring columns are of a known composition, the only significant variable is the depth position of the defect within the sample. Probe focus is set to Scherzer defocus, which on the sample is the focus that gives the sharpest image when imaging the bulk crystal. Conventional simulation studies tend to use only a single detector to image defects. Instead of imaging a single defect with a single ADF detector, here we take advantage of a laterally-displaced common defect *pair,* $In_{Zn}V_{Zn}$, of comparatively high and zero mass, as well as a multiple detector setup to find the most probable conditions for successfully measuring a defect's 3D-position.

Several authors (13–16) have shown how the STEM probe is prone to scatter forward and backward between neighbouring atomic columns in an oscillating fashion. Hwang et al. (14) showed by STEM simulation that Cs-corrected instruments are particularly prone to such scatter. In their $SrTiO_3$ example, the first maximum of probe intensity occurs at 0.8nm, after which a significant fraction of the intensity scatters to neighbouring columns. After this maxima, direct and intuitive correlation of intensities becomes much more complex and increases uncertainty in atom-counting. Hence, atom-counting techniques are not suitable when attempting to detect vacancy or substitution cases with very few atoms. Instead, the present paper presents a solution in the form of STEM image simulation in combination with numerical analysis and pattern recognition.

## 2. Methodology

### 2.1. Simulation details

The Prismatic (8,10) STEM simulation software were used to perform the simulations. Prismatic can utilize both the CPU and GPU, and always provides the full range of possible detector angles as output, with a given step, up to the maximum collection angle given by the potential spacing. The simulation computers were three servers with 28-core Intel Xeon CPU, 128 GB of ram and four Nvidia GeForce RTX 2080TI graphics cards. With Prismatic we were able to quickly simulate images with and without a defect with a small (1 mrad) step in acceptance angle, in order to understand over which acceptance angle regime we should be measuring the defect.

In order to ensure accurate simulation, the potential pixel size *p* was set to 5pm. This was smaller than the 8pm spatial resolution (probe spacing) of the simulated electron beam, ensuring that the atoms did not appear pixelated. Since prismatic additionally uses an anti-aliasing aperture 0.5 times the maximum scattering angle, and the accelerating voltage was set to 300kV, the effective maximum scattering angle was 98 mrad, according to equation 1.

$$\max scattering\ angle = \frac{\lambda q}{2}$$

where *q = 1/(2p)* is the reciprocal-space pixel size of the real space potential grid spacing *p* and lambda is the relativistic wavelength of the incoming electron. A final restriction is the angular resolution of the beam on the sample. Egerton (17) writes that the ratio of the incident beam semi-angle $\alpha$ should be at least ten times the angular resolution. This condition is described by equation 2,

$$\frac{\alpha}{angular\ resolution} = \frac{\alpha x}{\lambda} \geq 10$$

where x is the shortest lateral sample length, 72Å for our models. With the large cell size, the PRISM interpolation factor was set to 4, which gave fast simulation speeds with accurate results. For our simulations with a $\alpha$=20mrad (and a PRISM alpha limit of 22mrad), the ratio of $\alpha$ to angular resolution was 50.8. Simulations were performed with spherical aberration of 0.001mm and the beam focused on the top surface of the samples. It should be noted that in Prismatic, the beam hits the atoms with highest value of z first, and atoms at z=0 last. This is the opposite convention of most other simulation software. Chromatic aberration due to 0.9 eV energy spread of the incoming electron was included by a defocus series of five steps, equally spaced by the standard deviation of 48Å, as seen in Figure 1a.

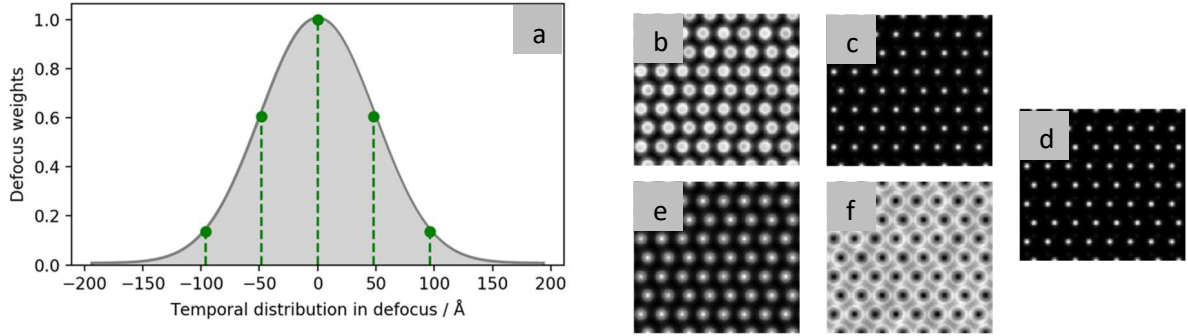

*Figure 1. a) Defocus series distribution according to an energy spread of 0.9eV. Chromatic aberration is approximated by weighted averaging of the resultant STEM images from five defocus values chosen 1 standard deviation (48Å) apart. The resulting images on bulk ZnO at room-temperature, are shown in b-e, for defocus of values from -96 through +96Å.*

## 2.2. Creating the defect models

As a starting point for modelling, hexagonal unit cells of ZnO were transformed into orthogonal cells. The orthogonal angles of these cells made it easier to construct and cut samples. Supercells of 3x3x2 unit cells of orthogonal ZnO were then constructed, with dimensions a=9.73Å, b=11.24Å, c=10.38Å, and rotated to give an orientation with the beam-direction along the 100 zone axis. Defect cells were built based on these supercells. The *static* defect model was modelled by removal of a single Zn and In substitution of a neighbouring Zn on the 110 plane. This *static* model was then relaxed by density functional theory (DFT) to produce the *relaxed* defect model. Larger models for STEM simulation were constructed using the Atomic Simulation Environment (18) Python software.

DFT calculations were performed using the Heyd-Scuseria-Ernzerhof (HSE) (19,20) hybrid functional and the projector augmented wave method (21–23), as implemented in the VASP code (21,23). The screening parameter was fixed to the standard value to $\omega = 0.2$ Å$^{-1}$, and the fraction of screened Hartree-Fock exchange adjusted to $\alpha = 0.375$ (24). Defect calculations were performed with the 3x3x2 supercell by keeping the lattice parameters fixed and relaxing all atomic positions until the residual forces were reduced to less than 5 meV/Å. The cut-off energy for the plane-wave basis set was set to 500 eV, and a special off-Γ *k*-point at $k = (¼,¼,¼)$ was used for integrations over the Brillouin zone. The maximum radial displacements from the static model in three-dimensions were 0.114Å and 0.311Å for Zn and O, respectively. Of the three-dimensional displacement, most was in the lateral direction, perpendicular to the beam direction. The maximum lateral displacements were 0.104Å and 0.311Å.

6 supercells of pristine ZnO were stacked along the beam direction. To model the defect as a function of depth, the defect supercell was inserted into the bulk at increasing depths, with steps of 1/3 of the defect cell depth. Total sample thickness was approximately 3.2nm. Once the defect structure was built, pristine ZnO was stacked laterally to give sample width and heights of 50 nm. Figure 2 shows the structure of the relaxed supercell, with labels indicating the defect-containing columns.

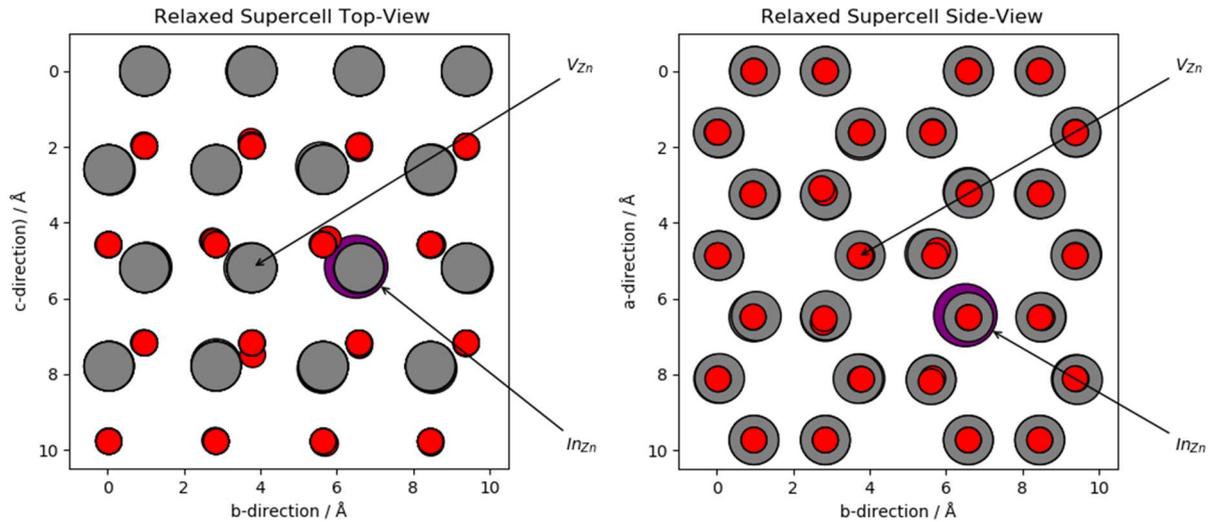

*Figure 2. Top-view (left, along 100) and side-view (right, along 001) representation of the defect supercell. Labels indicate defect-containing columns.*

While for larger models, 20-40 Frozen Phonon (FP) configurations are often enough to produce a realistic simulation, the thin specimen model demands a much higher number of FP configurations. To estimate the necessary number of FPs, convergence testing was performed on bulk ZnO. Simulation was performed on a region of interest (ROI) containing 30 columns of ZnO of the same thickness as the defect models. Each frozen phonon configuration was generated by random lateral displacement of the atom according to its Debye Waller factor (shown in Table 1) by the Prismatic software.

Figure 3 shows the intensity for 42 atomic columns, centred around the static defect model. The right-hand graph shows the Voronoi integrated intensity as a function of increasing number of FP. For these thin specimens 100 phonon configurations, laterally displaced, were shown to be sufficient to minimize error. Since each image is composed of the weighted average of a defocus series with five values, each image consists of 500 frozen phonon iterations.

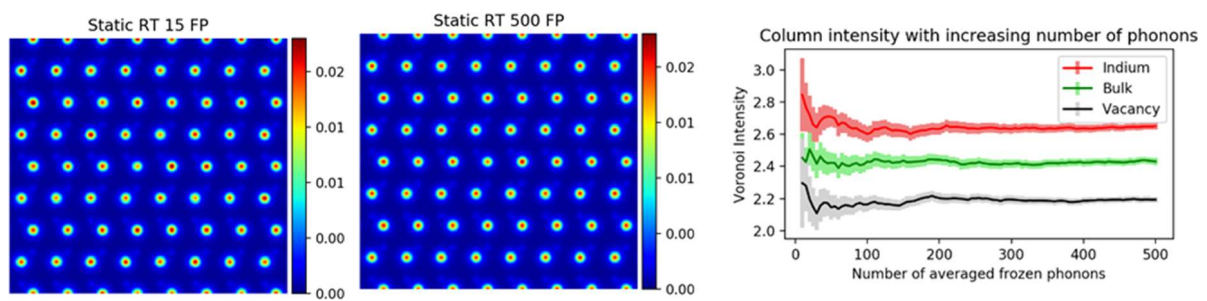

*Figure 3. Left: HAADF simulation of the static defect model at room temperature at depth 1 with 15 and 500 frozen phonons (FP). Right: Mean of Voronoi intensity by with increasing number of frozen phonons on the static defect at depth 1. The "Bulk" value here is taken as one of the edge columns on the image furthest from the defect. Error bars are one standard error.*

## 2.3. Analysis details

Each frozen phonon image resulted in a spectrum image with a file size of 28 MB. The 30 models simulated with 500 phonons totalled over 400 GB of data. To load such data, the lazy loading capability of the Hyperspy (25) open-source Python software was employed, which made it able to manipulate large data without loading it all into memory simultaneously. In order to measure the intensity of a given column, the Absolute Integration feature of the Atomap (26) open-source Python software was used. The integrator, which has previously (27,28) been used for numerical studies on STEM images is a method based on Voronoi cell integration. The smallest lateral distance between the Zn columns of the simulated model is 2.7Å. This is larger than the 2Å minimum distance recommended by De Backer et al. (29) for analysis like Voronoi cell integration that does not take peak overlap (such as could be fitted by two gaussians) into account. In its Atomap implementation, Voronoi integration can be limited to within circles of a given radii, in order to prevent edge-effects from interfering with the region of interest. In the present work, simulation was performed on an area large enough to facilitate full Voronoi integration. The Voronoi cells from columns at the edge of the images were removed as their areas differed due to interaction with the image border. Automatic removal of such edge effects, along with significant performance increases have been merged into the Atomap software by the authors.

## 2.4. Choosing optimal acceptance angles

STEM instruments have annular detectors that are fixed at a certain distance, and hence angle, from the specimen. Instruments that have post-specimen lenses can control the camera length from the specimen, which in turn affects the inner and outer acceptance angle for a given annular detector. On high-end STEM instruments, it is not unusual to have three or four annular detectors covering several ranges of acceptance angles, which can all simultaneously record images. These, combined with the ability to change camera length give rise to a high number of possible acceptance angles. If the microscope allows it, this can be further enhanced by shadowing one detector by another, thereby reducing the outer acceptance angle on the detector furthest from the specimen.

To determine the acceptance angles giving the highest contrast between pristine and defect ZnO regardless of defect depth position, bulk and static models were simulated across acceptance angles from 0 to 100 mrad in steps of 1 mrad. Then, for each equivalent pixel in the resulting two image stacks, the Michelson contrast was computed. The contrast value of all pixels for a given contrasted acceptance angle was then summed. This procedure was done for defect-complexes sat at increasing depth in the sample (depth position is further discussed later). The same procedure was applied to the bulk and relaxed models, giving slightly different contrast values. The contrasts were offset relative to each other and are displayed in Figure 4. It is shown that high contrast is found from approximately 35 mrad and upwards. At the medium-angle annular dark-field (MAADF), the total contrast is shown to vary strongly with depth. At the high-angle annular dark-field (65 mrad and upwards), the total contrast appears nearly unchanging as a function of depth.

Two ranges of acceptance angles were chosen for further study. A high-contrast HAADF region from 65-95 mrad was chosen to maximize the probability of finding a defect region, and a MAADF region of varying contrast from 35-45 mrad was chosen to distinguish the depth-position of the defect-complex within the sample.

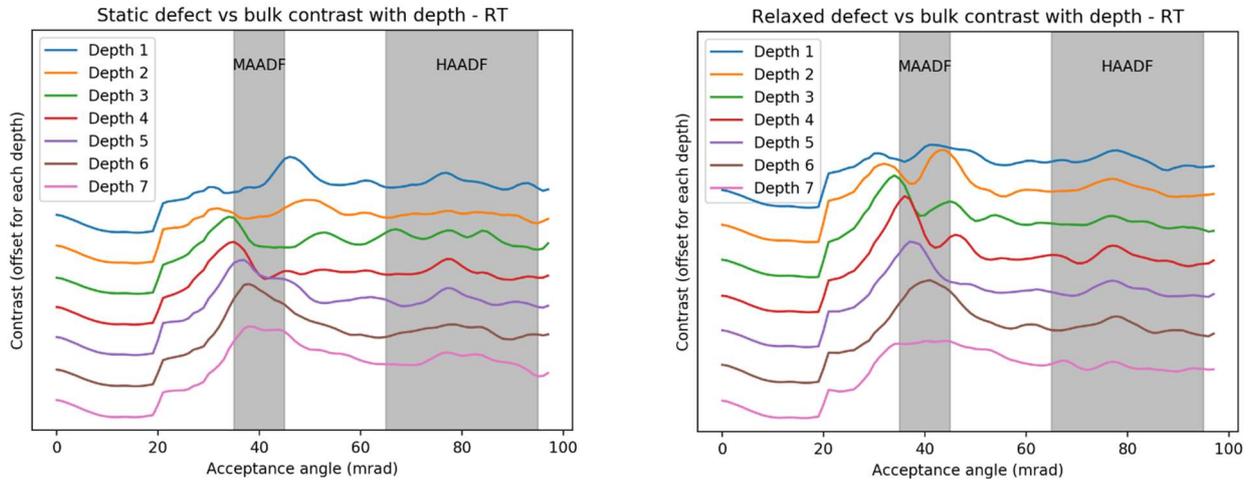

*Figure 4. Michelson contrast comparing (left) bulk and static and (right) bulk and relaxed defect ZnO as a function of acceptance angle and defect depth position. MAADF and HAADF ranges of acceptance angles highlighted in grey are chosen for later image comparisons.*

Acceptance angles in the range 65-95 mrad are typical of the HAADF detector on the microscope. With the chosen convergence angle of 20 mrad, acceptance angles in the range 35-45 mrad constitute are detectable on an ADF detector placed in this range, possibly shadowed by another detector. MAADF images typically contain a mix of diffraction contrast and atomic number contrast, which are sensible contrast mechanisms for the material system we are investigating. With realistic acceptance angles to construct the STEM images, the differences between STEM simulation on the static and relaxed defect structures were calculated.

## 3. Results

### 3.1. Comparing DFT-relaxed defect cell with static defect cell

The effect and importance of DFT-relaxation of the static defect cell is difficult to see by eye on the standard STEM image, but the differences become clear by numerical analysis. Figure 5 shows the differences in intensity of the static (top) and relaxed (bottom) cells, showing both the raw MAADF STEM simulation but also the integrated Voronoi image. The intensities summed in the Voronoi images have been normalised by the equivalent intensity for pure ZnO. The defect-containing columns stand out in accordance with the typically referred-to notion of linear dependence on number of atoms and a $Z^2$-dependence for HAADF images (30). There is no distinct change in the intensity on the neighbouring columns around the defect.

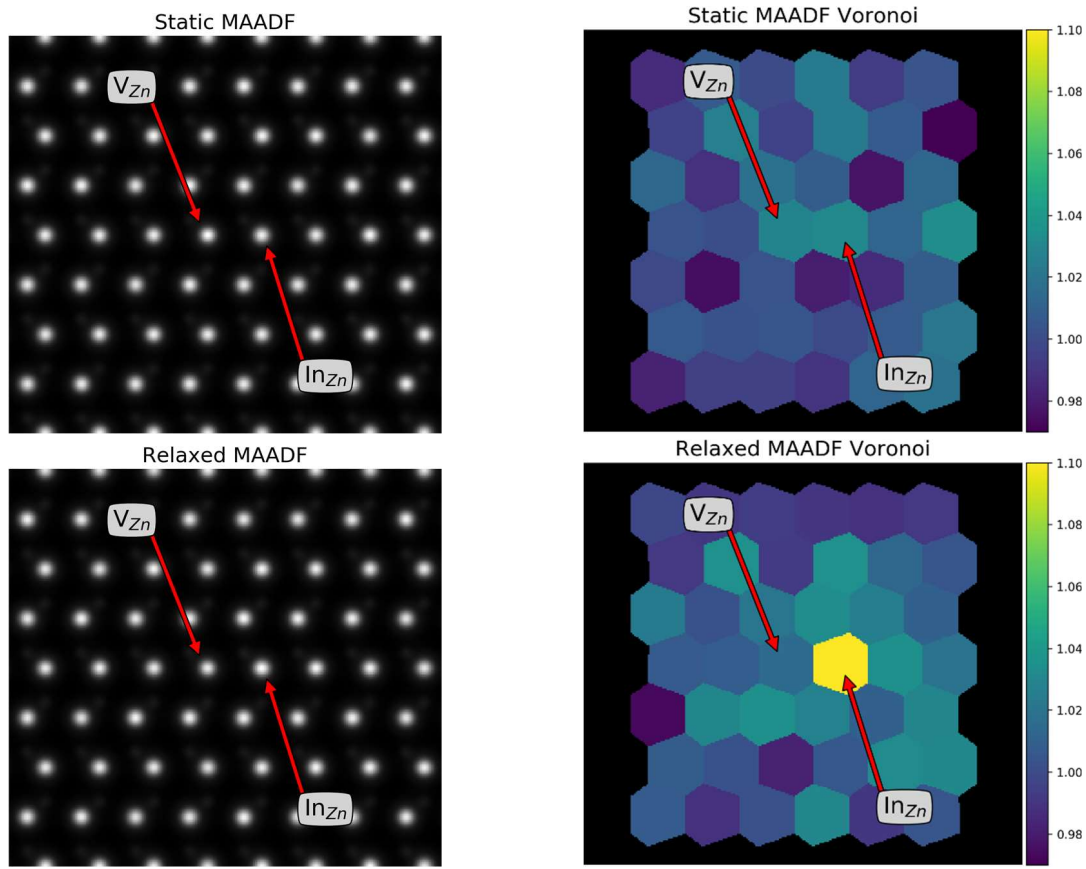

*Figure 5. Conversion of simulated HAADF image of the model of the In-V defect-complex in ZnO (left) to the intensity image by Voronoi-integration (right), taken from a simulation with the defect at position 7. Top: Static model. Bottom: Relaxed model. The $In_{Zn}$ and $V_{Zn}$-containing columns are highlighted.*

## 3.2. Effect of defect depth position on contrast

To further investigate the intensity variation of the defect, the defect was modelled as a function of depth. To preserve relaxation around the defect, the first depth position containing the indium defect was the fifth layer. Then, the defect supercell was stepped down throughout the bulk model until the indium defect reached the second layer from the bottom of the model. This is shown on the right-hand schematic in Figure 6.

After Voronoi-integration, the intensities of the two defect-containing columns were normalised by the bulk mean and plotted in Figure 6, as a function of depth. The HAADF signal shows typical Z-contrast, displaying intensities of the expected trend of $In_{Zn}$ > Bulk > $V_{Zn}$ due to the atomic number of the atoms and number of atoms present in the column. The indium-containing column (hereafter "In-column") is on average 26% brighter than the vacancy-column (V-column) on both the static and relaxed model, with small variation of the intensities with defect depth position. This is beneficial for locating the defect laterally on the sample but is of no value in determining its depth-position. There is little difference between the static and relaxed models for the HAADF signal.

The MAADF signal shows strong variation in intensity with depth, particularly for the $V_{Zn}$-containing column. When near the top of the sample, the defect-complex (particularly the vacancy column) is virtually indistinguishable from the average bulk intensity in the MAADF image, whilst intense in the HAADF. However, as the defect moves down through the sample it increases significantly in MAADF contrast. When located at the penultimate position, the $In_{Zn}$ column is 52% and 60% brighter than the $V_{Zn}$ column, and 16% and 23% brighter than the bulk atom for the relaxed and static models, respectively. This variation in contrast makes it possible to determine the approximate depth-position of the defect.

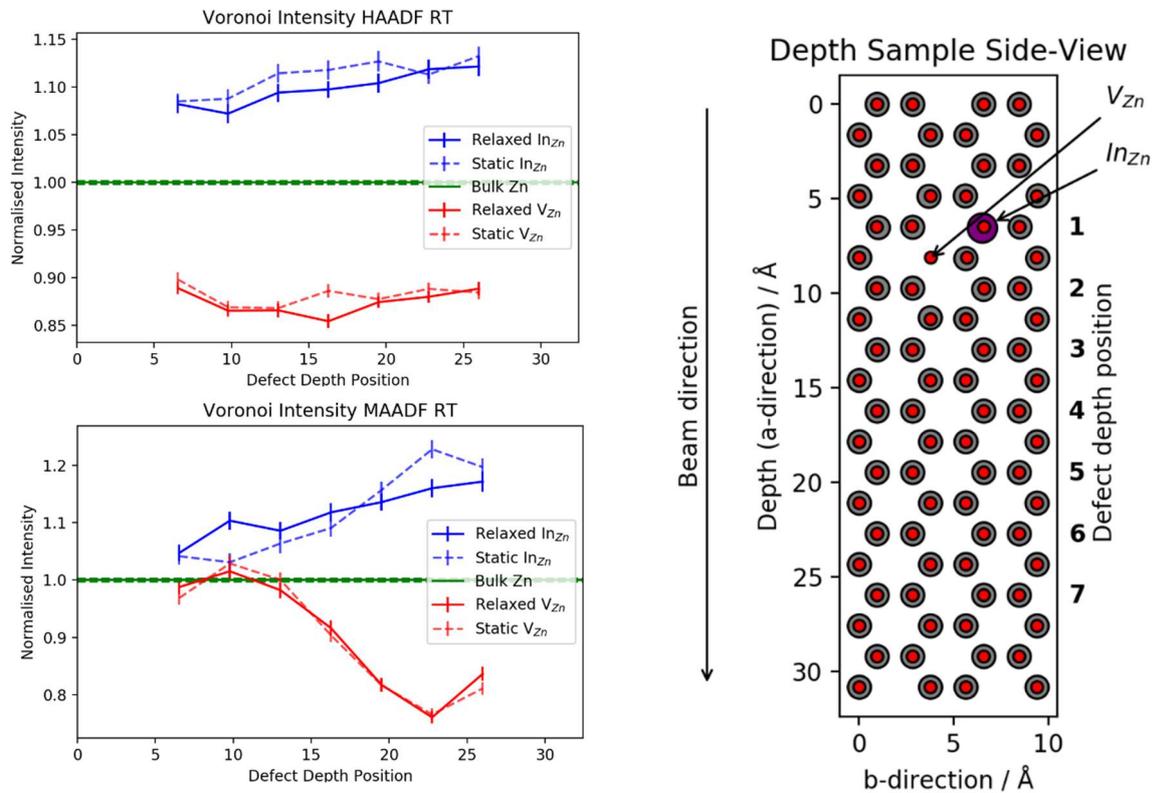

Figure 6. Top: Integrated HAADF intensity over the atomic columns as a function of depth position of the defect complex for a static and DFT-relaxed model with 500 frozen phonon configurations at room-temperature. Bottom: Same as top but for the MAADF signal. Right: Schematic of a cross-section of the sample showing the defect position. The right-hand axis shows the position of the $In_{Zn}$ in the defect-complex as it is stepped throughout the sample.

The change in defect intensity is even more clearly visualised by looking at the Voronoi image as a function of depth. Figure 7 shows the Voronoi intensity for HAADF (top) and MAADF (bottom) intensity normalised for the bulk mean. Here it is clear that the MAADF pattern created by the defect becomes stronger with increasing depth, with the In-column column becoming visible first.

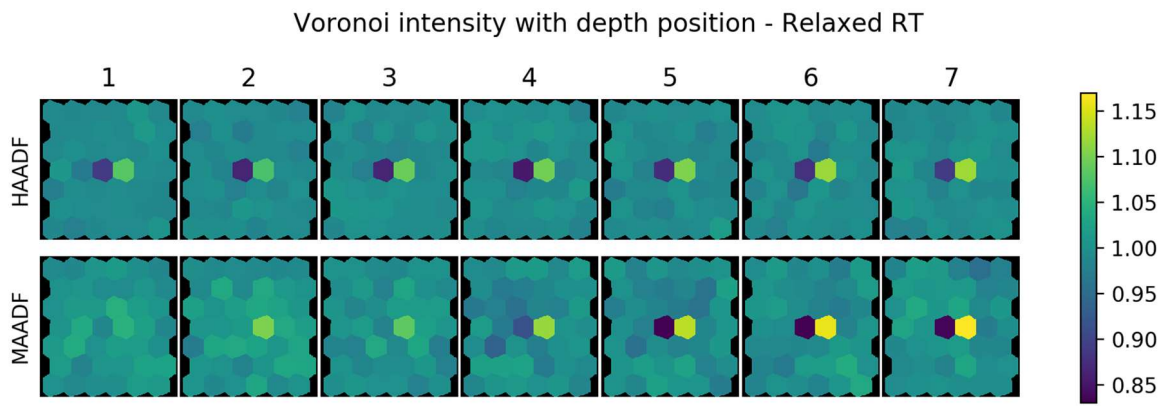

*Figure 7. Voronoi intensity on the Relaxed model at room temperature with increasing depth. HAADF and MAADF images have the same colour bar.*

## 3.3. Effect of temperature on sample

Modern STEM instruments often have the addition of cooling holders to cool the sample down using liquid nitrogen, typically to about 100K. To investigate the effect of cooling on the STEM intensity, we repeated the defect depth-study with Debye-Waller factors for ZnO at 100K instead of 300K. A temperature-dependent model for the Debye-Waller factor of Zn and O in ZnO is given by (31), and is plotted in Figure 8. The Debye-Waller values chosen for simulation are shown in Table 1. The value for the In defect was set equal to the Zn value, since no Debye-Waller value was available for the In defect and the Zn and O values were very close despite their difference in atomic number.

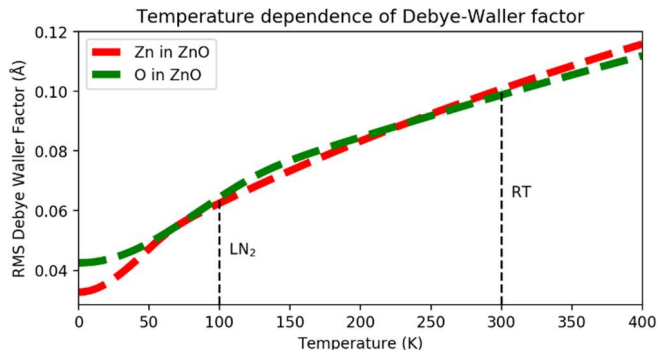

| Temperature (K) | RMS Debye-Waller Factor (Å) | |
|---|---|---|
|  | Zn | O |
| 300 | 0.101 | 0.099 |
| 100 | 0.062 | 0.064 |

*Figure 8. Temperature-dependence of Zn and O in ZnO based on a model by (31).*

*Table 1. Values of the root-mean-square Debye-Waller factor at 300 and 100K from (31)*

The Voronoi-integrated intensities of the same columns as shown in Figure 6 are shown in Figure 9. The liquid-nitrogen HAADF results are similar to the room-temperature results, but with an overall increase in intensity of the defect columns relative to the bulk. This shifts the In-column up from the bulk average but reduces the intensity gap between the bulk and the V-column. For both temperatures, the In-column shows an increasing intensity with depth position, but this behaviour is not seen for the V-column. The MAADF intensity is greatly changed. For the first three depth positions, the V-column is brighter than the In-column. At deeper positions, the defect increases strongly in contrast relative to the room-temperature simulation. At its highest, the In-column is 85% brighter than its counterpart.

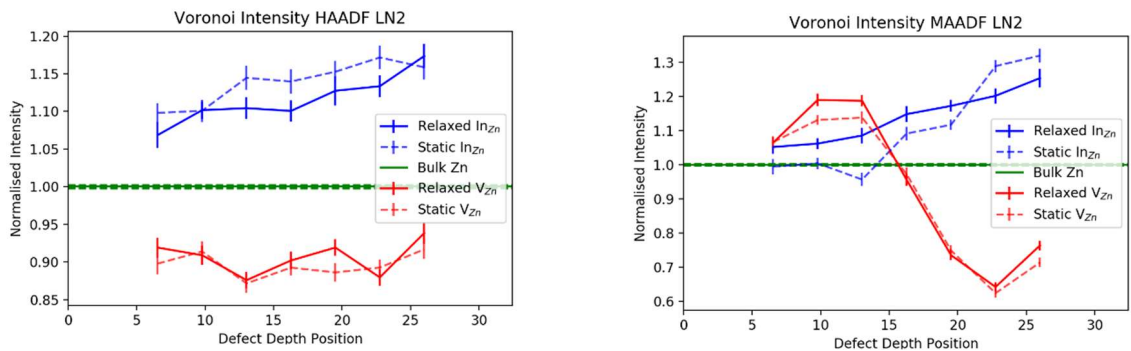

*Figure 9. Left: Integrated HAADF intensity over the atomic columns as a function of depth position of the defect complex for a static and DFT-relaxed model with 500 frozen phonon configurations at liquid nitrogen temperature. Right: Same as left but for the MAADF signal.*

The DFT-relaxation has approximately the same impact on the defect columns at liquid nitrogen temperatures as at room-temperature but reduces the HAADF In-column intensity more at the lower temperature. The V-column experiences only a small relaxation effect on the MAADF image. The second and third depth position are brighter than expected by the static model, but for all other depths the vacancy appears nearly identical in both cases. Noticeably, there is a contrast reversal at depth-position 4, where the V-column becomes darker than its counterpart.

There is a significant change in intensity on the columns *surrounding* the defect on the relaxed model at liquid nitrogen temperature, as seen in Figure 10. This effect is not seen on the static model nor seen on the simulations done at room-temperature, but only visible on the Relaxed model MAADF image at liquid nitrogen temperature.

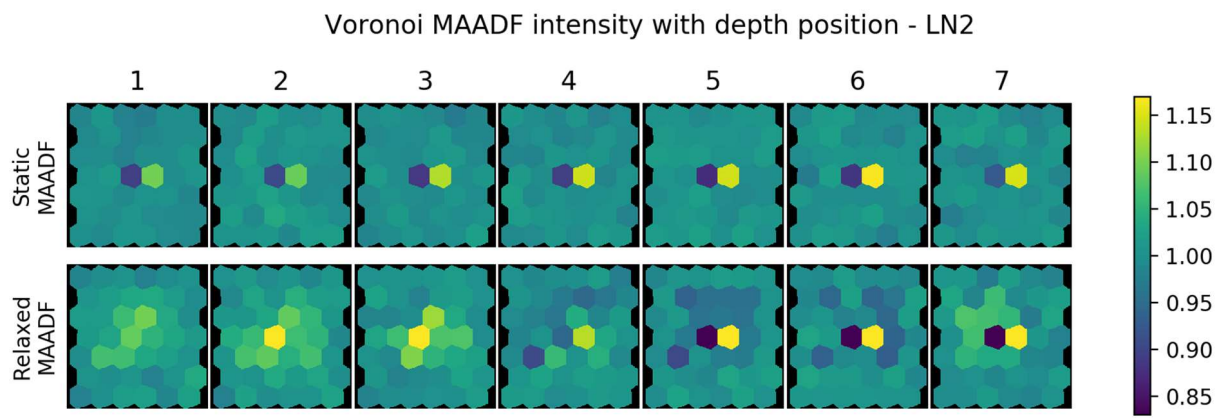

*Figure 10. Voronoi MAADF intensity on the Static (top) and Relaxed (bottom) model at liquid nitrogen temperature with increasing depth. Both sets of images have the same have the same colour bar. The effect on surrounding columns is only seen for the relaxed defect.*

## 4. Discussion

Voronoi cell integration of high-resolution STEM images has been shown to be an efficient method to simplify identification of changes to column intensity without resorting to curve fitting or maxima approaches. While the method yields less information (such as the width of the column) than conventional curve-fitting algorithms, it is much faster and can be performed on very large datasets in a matter of seconds. However, it is important to note that it requires sufficient spatial resolution (80 pm was determined to be enough for this work) to not introduce sampling artefacts, and datasets acquired experimentally must be done so with long enough exposure to minimize noise.

The HAADF intensity is mostly independent with the depth-position of the defect. This is in agreement with previous results (5). However, for the In-column, there is an increased intensity for the last three positions as the defect approaches the bottom of the sample. Zhang et al. (32) provide a likely explanation of the increase by electron channelling. They argue that as the incident probe channels along a column, atoms deeper in the foil see a more focused probe and consequently scatter to higher angles due to the closer proximity of the probe to the nucleus. This explains the brighter column intensity of the heavy In defect with depth. The V-column is dimmest when the vacancy is in the middle of the sample, becoming slightly brighter at the edges. This is likely due to the vacancy disrupting the effect of the channelling-focused probe, causing scatter to lower angles instead.

The MAADF intensity has been found to vary significantly with depth position, and alongside the lateral position gained from the HAADF intensity, allows for a full three-dimensional measurement of the defect position. This work complements similar analysis done by Johnson et al. (5) on single point-defects. The depth variation is likely due to a shift in scattering angle as the defect proceeds deeper into the sample. The peaks in the MAADF range on Figure 4 show a clear trend to the right with depth position, confirming this suspicion. It is possible that the same effect described by Zhang et al. (32) for the HAADF regime is present here.

The DFT relaxation affects the In HAADF intensity in a nearly consistent manner, significantly reducing the intensity. This reduction can be attributed to the lateral shift of the indium atom with respect to the channelling-focused probe. Since the probe no longer targets the centre of the atom, this reduces the likelihood of the high-angle electron-nucleus scattering associated with the HAADF imaging.

The measurements at liquid nitrogen temperature show a bigger difference between the relaxed and static models than for the measurements at room-temperature. The biggest change is found on the In column. This is likely due to the Debye-Waller factors of Zn and O being much higher at room temperature than at liquid nitrogen temperatures, which results in a smearing of the intensity across acceptance angles. The smaller change on the vacancy column is expected due to the lack of a Debye-Waller effect associated with the vacancy.

The change in MAADF intensity of the neighbouring columns at liquid-nitrogen temperatures due to defect relaxation is a novel important observation. The intensity change makes it easier to spot such defects, but also implies that other types of defects may cause significant intensity changes on several columns at liquid nitrogen temperatures. Hence, depending on the type of defect one wishes to measure, it may be either an advantage or a disadvantage to image at liquid nitrogen temperatures. More types of defects would need to be investigated to determine whether this is a general phenomenon.

An exciting avenue for single-defect detection is template matching (33,34). By manipulating the integrated intensities from the simulation, a "defect-complex" pattern could be created and fit across large STEM images to find the position of defects in an objective manner. If a simulated static image is used to "hunt" for defects by attempting to pattern match across a large STEM image, it may find such a defect but estimate an incorrect z-position in the sample. Template matching based on simulated data is in its infancy, but here we report an important consideration for future work.

## 5. Conclusions

The three-dimensional position information of a defect-complex in ZnO has been shown to be possible to determine by STEM. A multi-detector configuration with specific acceptance angles is shown to be particularly helpful to maximise contrast and the chance of detecting the defect. The particular angles are shown to be computable by a straightforward contrast calculation on simulated data. While the lateral position could possibly be determined via conventional "atom-counting" techniques on HAADF images, we have shown that STEM simulation is essential to determine the depth-position within the sample from experimental STEM images. The impact of a DFT-relaxed simulation model is shown to be particularly important for simulations with low Debye-Waller factors, either due to the elements involved or temperature considerations. Particularly, for simulating template-matching templates, relaxation is crucial. With the high energy-resolution present in the newest generation monochromated STEM instruments, low noise in energy filter cameras, and true measurements of the position of a defect described in this paper, it is possible to begin investigating the optical and electronic properties of such defects by EELS measurements.

## 6. Acknowledgements

The authors are grateful to Rolf Erni, Ivan Lobato, Alan Pryor and Colin Ophus for valuable discussion and help with simulation software.

The authors would like to acknowledge support from the Research council of Norway through the Norwegian Center for Transmission Electron Microscopy, NORTEM (197405/F50), the Norwegian Micro- and Nano-Fabrication Facility, NorFab (197411/V30), and the FriPRO Toppforsk project FUNDAMeNT (no. 251131). DFT computations were performed on resources provided by UNINETT Sigma2 - the National Infrastructure for High Performance Computing and Data Storage in Norway. STEM simulation was performed on hardware generously provided by the University of Oslo AI Hub.

1. Ozgur Ü, Hofstetter D, Morkoç H. ZnO devices and applications: A review of current status and future prospects. Proc IEEE. 2010;98(7):1255–68.

2. Nomoto J ichi, Konagai M, Okada K, Ito T, Miyata T, Minami T. Comparative study of resistivity characteristics between transparent conducting AZO and GZO thin films for use at high temperatures. Thin Solid Films [Internet]. 2010;518(11):2937–40. Available from: http://dx.doi.org/10.1016/j.tsf.2009.10.134

3. Minami T. Present status of transparent conducting oxide thin-film development for Indium-Tin-Oxide (ITO) substitutes. Thin Solid Films. 2008;516(17):5822–8.

4. Minami T, Sato H, Nanto H, Takata S. Group III impurity doped zinc oxide thin films prepared by RF magnetron sputtering. Jpn J Appl Phys. 1985;24(10):L781–4.

5. Johnson JM, Im S, Windl W, Hwang J. Three-dimensional imaging of individual point defects using selective detection angles in annular dark field scanning transmission electron microscopy. Ultramicroscopy [Internet]. 2017;172(June 2016):17–29. Available from: http://dx.doi.org/10.1016/j.ultramic.2016.10.007

6. Lobato I, van Aert S, Verbeeck J. Progress and new advances in simulating electron microscopy datasets using MULTEM. Ultramicroscopy [Internet]. 2016;168:17–27. Available from: http://dx.doi.org/10.1016/j.ultramic.2016.06.003

7. Pryor A, Ophus C, Miao J. A Streaming Multi-GPU Implementation of Image Simulation Algorithms for Scanning Transmission Electron Microscopy. Adv Struct Chem Imaging [Internet]. 2017; Available from: http://arxiv.org/abs/1706.08563

8. Allen LJ, D'Alfonso AJ, Findlay SD. Modelling the inelastic scattering of fast electrons. Ultramicroscopy [Internet]. 2015;151:11–22. Available from: http://dx.doi.org/10.1016/j.ultramic.2014.10.011

9. Ophus C. A fast image simulation algorithm for scanning transmission electron microscopy.

Adv Struct Chem Imaging [Internet]. 2017;3(1):13. Available from: http://ascimaging.springeropen.com/articles/10.1186/s40679-017-0046-1

10. Lobato I, Van Dyck D. MULTEM: A new multislice program to perform accurate and fast electron diffraction and imaging simulations using Graphics Processing Units with CUDA. Ultramicroscopy [Internet]. 2015;156:9–17. Available from: http://dx.doi.org/10.1016/j.ultramic.2015.04.016

11. Oelerich JO, Duschek L, Belz J, Beyer A, Baranovskii SD, Volz K. STEMsalabim: A high-performance computing cluster friendly code for scanning transmission electron microscopy image simulations of thin specimens. Ultramicroscopy [Internet]. 2017;177:91–6. Available from: http://dx.doi.org/10.1016/j.ultramic.2017.03.010

12. Mittal A, Mkhoyan KA. Limits in detecting an individual dopant atom embedded in a crystal. Ultramicroscopy [Internet]. 2011;111(8):1101–10. Available from: http://dx.doi.org/10.1016/j.ultramic.2011.03.002

13. Hwang J, Zhang JY, D'Alfonso AJ, Allen LJ, Stemmer S. Three-dimensional imaging of individual dopant atoms in SrTiO3. Phys Rev Lett. 2013;111(26):1–5.

14. Ishikawa R, Lupini AR, Findlay SD, Taniguchi T, Pennycook SJ. Three-dimensional location of a single dopant with atomic precision by aberration-corrected scanning transmission electron microscopy. Nano Lett. 2014;14(4):1903–8.

15. Voyles PM, Muller DA, Grazul JL, Citrin PH, Gossmann HJL. Atomic-scale imaging of individual dopant atoms and clusters in highly n-type bulk Si. Nature. 2002;416(6883):826–9.

16. Egerton R. Electron Energy-Loss Spectroscopy in the Electron Microscope [Internet]. 3rd ed. Book. Springer US; 2011. 491 p. Available from: http://link.springer.com/10.1007/978-1-4419-9583-4%5Cnhttp://www.springerlink.com/index/10.1007/978-1-4419-9583-4

17. Larsen AH, Jens J, Blomqvist J. The atomic simulation environment — a Python library for working with atoms. 2017;2000.

18. Krukau A V., Vydrov OA, Izmaylov AF, Scuseria GE. Influence of the exchange screening parameter on the performance of screened hybrid functionals. J Chem Phys. 2006;125(22):0–5.

19. Perdew JP, Burke K, Ernzerhof M. Generalized Gradient Approximation Made Simple. Phys Rev Lett. 1996;77(18):3865–8.

20. Blöchl PE. Projector augmented-wave method. Phys Rev B. 1994;50(24):17953–79.

21. Kresse G, Hafner J. Ab initio molecular dynamics for liquid metals. Phys Rev B. 1993;47(1):558–61.

22. Kresse G, Joubert D. From ultrasoft pseudopotentials to the projector augmented-wave method. Phys Rev B. 1999;59(3):1758–75.

23. Kresse G, Furthmuller J. Efficient iterative schemes for ab initio total-energy calculations using a plane-wave basis set. Phys Rev B. 1996;54(16):11169–86.

24. Oba F, Togo A, Tanaka I, Paier J, Kresse G. Defect energetics in ZnO: A hybrid Hartree-Fock density functional study. Phys Rev B - Condens Matter Mater Phys. 2008;77(24):3–8.

25. de la Peña F, Fauske VT, Burdet P, Ostasevicius T, Sarahan M, Nord M, et al. Hyperspy

software package [Internet]. 2018. Available from: http://dx.doi.org/10.5281/zenodo.54004

26. Nord M, Vullum PE, MacLaren I, Tybell T, Holmestad R. Atomap: a new software tool for the automated analysis of atomic resolution images using two-dimensional Gaussian fitting. Adv Struct Chem Imaging. 2017;3(1):9.

27. MacArthur KE, Brown HG, Findlay SD, Allen LJ. Probing the effect of electron channelling on atomic resolution energy dispersive X-ray quantification. Ultramicroscopy. 2017;182:264–75.

28. Jones L, Macarthur KE, Fauske VT, Van Helvoort ATJ, Nellist PD. Rapid estimation of catalyst nanoparticle morphology and atomic-coordination by high-resolution Z-contrast electron microscopy. Nano Lett. 2014;14(11):6336–41.

29. De Backer A, van den Bos KHW, Van den Broek W, Sijbers J, Van Aert S. StatSTEM: An efficient approach for accurate and precise model-based quantification of atomic resolution electron microscopy images. Ultramicroscopy. 2016;171:104–16.

30. Williams DB, Carter CB. Transmission Electron Microscopy: A Textbook for Materials Science [Internet]. Springer; 2009. 820 p. (Cambridge library collection). Available from: https://books.google.co.uk/books?id=dXdrG39VtUoC

31. Schowalter M, Rosenauer A, Titantah JT, Lamoen D. Temperature-dependent Debye-Waller factors for semiconductors with the wurtzite-type structure. Acta Crystallogr Sect A Found Crystallogr. 2009;65(3):227–31.

32. Zhang JY, Hwang J, Isaac BJ, Stemmer S. Variable-angle high-angle annular dark-field imaging: Application to three-dimensional dopant atom profiling. Sci Rep [Internet]. 2015;5(July):1–10. Available from: http://dx.doi.org/10.1038/srep12419

33. Zuo JM, Shah AB, Kim H, Meng Y, Gao W, Rouviére JL. Lattice and strain analysis of atomic resolution Z-contrast images based on template matching. Ultramicroscopy [Internet]. 2014;136:50–60. Available from: http://dx.doi.org/10.1016/j.ultramic.2013.07.018

34. Ohwada M, Kimoto K, Mizoguchi T, Ebina Y, Sasaki T. Atomic structure of titania nanosheet with vacancies. Sci Rep. 2013;3(c):1–5.